\documentclass[a4paper,11pt]{article}
\usepackage{pos}

\title{Microscopic black holes as probes for quantum gravity}

\author*[a,b]{Samuel Kov\'a\v{c}ik}

\affiliation[a]{Faculty of Mathematics, Physics and Informatics, Comenius University in Bratislava, Bratislava, Slovakia}

\affiliation[b]{Department of Theoretical Physics and Astrophysics, Faculty of Science, Masaryk University, Brno, Czech Republic}

\emailAdd{samuel.kovacik@fmph.uniba.sk}
        
\abstract{One of the main goals of contemporary theoretical physics is to find the quantum theory of gravity. There are various working hypotheses, mostly operating in the regime of high-energy physics well above the reach of particle accelerators. So far, strong experimental or observational evidence to guide the theory is missing. A possible consequence of quantum gravity and quantum spacetime that is often discussed is the vacuum dispersion effect. In this paper, we consider a different line of quantum space phenomenology, the behaviour of microscopic black holes. Even though their exact nature is unknown, some of their features are very model-independent, allowing us to draw conclusions about their role in the current cosmological models.}

\FullConference{Corfu Summer Institute 2021 "School and Workshops on Elementary Particle Physics and Gravity"\\
  29 August - 9 October 2021\\
  Corfu, Greece}

\begin{document}

\maketitle

\section{Introduction}

Fundamental constants hint at important physical scales. We understand the importance of the speed of light, $c$, the gravitational constant, $G$, or the Planck constant, $\hbar$. The importance of their combinations, such as the Planck length
\begin{equation}
l_P = \sqrt{\frac{G \hbar}{c^3}} \approx 1.616255(18) \times 10^{-35} \mbox{m}
\end{equation}
is still being discussed. Many different perspectives view it as the minimal length of space(time) interval \cite{Hossenfelder:2012jw}. This length scale, or dual to it the Planck energy scale, is inaccessible for current particle accelerators. Phenomenology is therefore more focused on observational data of astrophysical origin. For example, gamma-ray bursts (GRB) travel through space for billions of years and minuscule effects of Planck-length physics can accumulate and manifest themselves as the vacuum dispersion \cite{Amelino-Camelia:1997ieq, Amelino-Camelia:1999jfz, Amelino-Camelia:2009imt,Sarkar:2002mg, Amelino-Camelia:2016ohi}.

Objects that can be sensitive to the existence of minimal length scale are black holes, as in the current general-relativity description their mass resides in a point-like singularity. The existence of a minimal length scale should lead to the absence of point-like objects, considerable research has been done on some aspects of regular black holes \cite{Dymnikova:1992ux, Bogojevic:1998ma, Wang:2009ay}. In this line of research, the singular matter distribution is replaced with a blurred one---for example described by the Gaussian distribution.

The solution to Einstein field equations with such matter distribution is then found and analysed. Mostly, the effects are negligible for macroscopic black holes of astrophysical origin. They are, however, important for microscopic black holes whose size---that is the Schwarzschild radius---approaches the Planck length. In many cases, the Hawking temperature of a regular black hole has an upper boundary and has a minimal Schwarzschild radius \cite{Nicolini:2005vd, Kovacik:2021qms}. When this minimum size is reached, the Hawking temperature vanishes. The resulting object has the cross-section of the order of the Planck area and has a considerable mass---of the order of the Planck mass. Therefore, it was suggested that such objects are viable dark matter candidates \cite{Chen:2002tu}.

What would be the origin of such small black holes, since astrophysical ones could not have evaporated this much during the current existence of the Universe? A possible source of \textit{small} black holes are the overdensities in primordial fluctuations. Their mass at the time of production has no lower boundary and their mass spectrum can be derived from various models of cosmological inflation \cite{Carr:2009jm, Khlopov:2008qy}.

Given our current knowledge, the existence of small black holes is neither guaranteed nor disproved. In this paper, we discuss their general properties, their role in the cosmological context, and outline future lines of research.
\section{Regular black holes}

Ordinary black holes are vacuum solutions of the Einstein field equations with curvature singularity at the origin. Regular black holes lack this singularity and have positive mass density, $\rho$, in some nonzero space volume. In principle, this mass distribution could be derived from the fundamental theory of quantum gravity, which is, however, currently unknown.

Regular black holes appear naturally in theories with quantum spaces that have a minimal length scale, denoted $\lambda$, which is expected to be of the Planck-length order. The construction goes as follows. First, one needs to find the minimal-length analogue of the point-particle Delta function; perhaps using coherent states as in the following references. For example, in the case of the 2D Moyal plane \cite{Moyal:1949sk, Amelino-Camelia:2008fcv,Doplicher:1994tu, Gayral:2003dm}, this yields $\rho (r) \sim e^{-\left(r/\lambda\right)^2}$, see \cite{Nicolini:2005vd}. In the case of 3D noncommutative rotationally invariant space \cite{Galikova:2013zca,KovacIk:2013vbk}, it gives $\rho (r) \sim e^{-\left(r/\lambda\right)}$, see \cite{Kovacik:2017tlg}. This matter density is then completed into the stress-energy tensor in a way so the solution to field equations can be of the Schwarzschild-like form, $g=\mbox{diag} \left(f(r),-f^{-1}(r), r^2, r^2 \sin \theta\right)$. This ansatz is then plugged into the field equations which can be expressed in the form of the following differential equation for the redshift factor $f(r)$
\begin{equation} \label{diff eq for f}
\frac{1 + f(r) + r f'(r)}{r^2} = 8 \pi \rho(r).
\end{equation}
The solution of this equation can be found in the form 
\begin{equation} \label{f int}
f(r) = -1 + \frac{2m}{r}\int \limits_0^r \tilde{\rho}(r') 4 \pi r'^2 dr',
\end{equation}

where $\tilde{\rho}(r)$ is a density normalised to unity, $\rho  (r) = m \tilde{\rho} (r)$. Regularity of this solution is obvious from the construction. 

How does it behave? Close to the origin is the function $f(r)$ growing from the value of $-1$. Far away from the origin, $r\gg \lambda$, it asymptotes the Schwarzschild solution, $-1 + \frac{2m}{r}$. It is obvious that $f(r)$ has a maximum for a finite and positive value of $r$ and its value at this point grows linearly with $m$. Let us, for simplicity, consider only monotonic matter densities $\rho(r)$. In that case, there is only one maximum and the value of $m$ determines whether is the maximum value positive, negative, or zero \cite{Kovacik:2021qms}. 

\begin{figure}[h] 
\centering
  \includegraphics[width=1.0\textwidth]{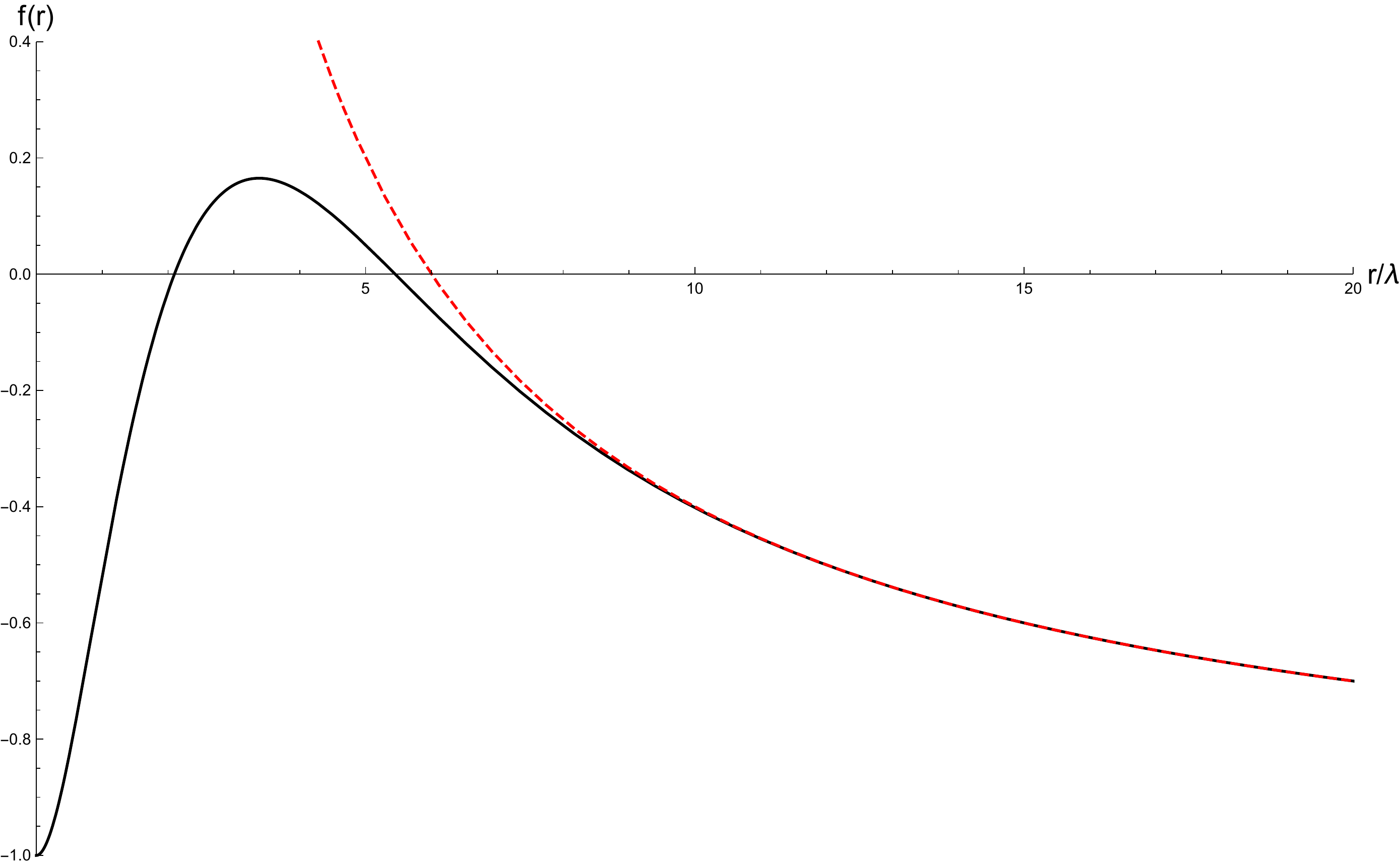}
    \caption{
Comparison of the redshift factor of a regular black hole (black solid line) and the Schwarzschild solution (red dashed line) with the same mass. In this case, the mass of the regular black hole is above the minimal mass, $m>m_0$, since there are two horizons. This solution is of the form $f(r) = -1 + \frac{2m}{r}\int \limits_0^r \tilde{\rho}(r') 4 \pi r'^2 dr'$, so it is obvious that decreasing the mass would lower the peak of this function and for a sufficiently small mass, there would no horizon at all.}
   \label{figure g}
\end{figure}

In case it is positive, since at limiting cases $r = 0$ and $r\rightarrow \infty$ is the function $f(r)$ negative, there have to be two horizons such that $f(r_\pm) = 0$. This follows from the continuity of $f(r)$. When the mass is very large, $m \gg \lambda$ (the Planck mass in our units), one of the horizons is close to the Schwarzschild value and one is close to the origin, see Figure \ref{figure g}. As the mass is being decreased, they move toward each other and merge at some specific value $m=m_0$. For $m<m_0$, there are no horizons---we have a blurred naked singularity. For $m=m_0$, the two horizons merge $r_+ = r_-$ and $f(r)$ reaches maximum at this point. From this follows that the derivative $f'(r)$ vanishes there and so does the Hawking temperature. 

\begin{figure}[h] 
\centering
  \includegraphics[width=1.0\textwidth]{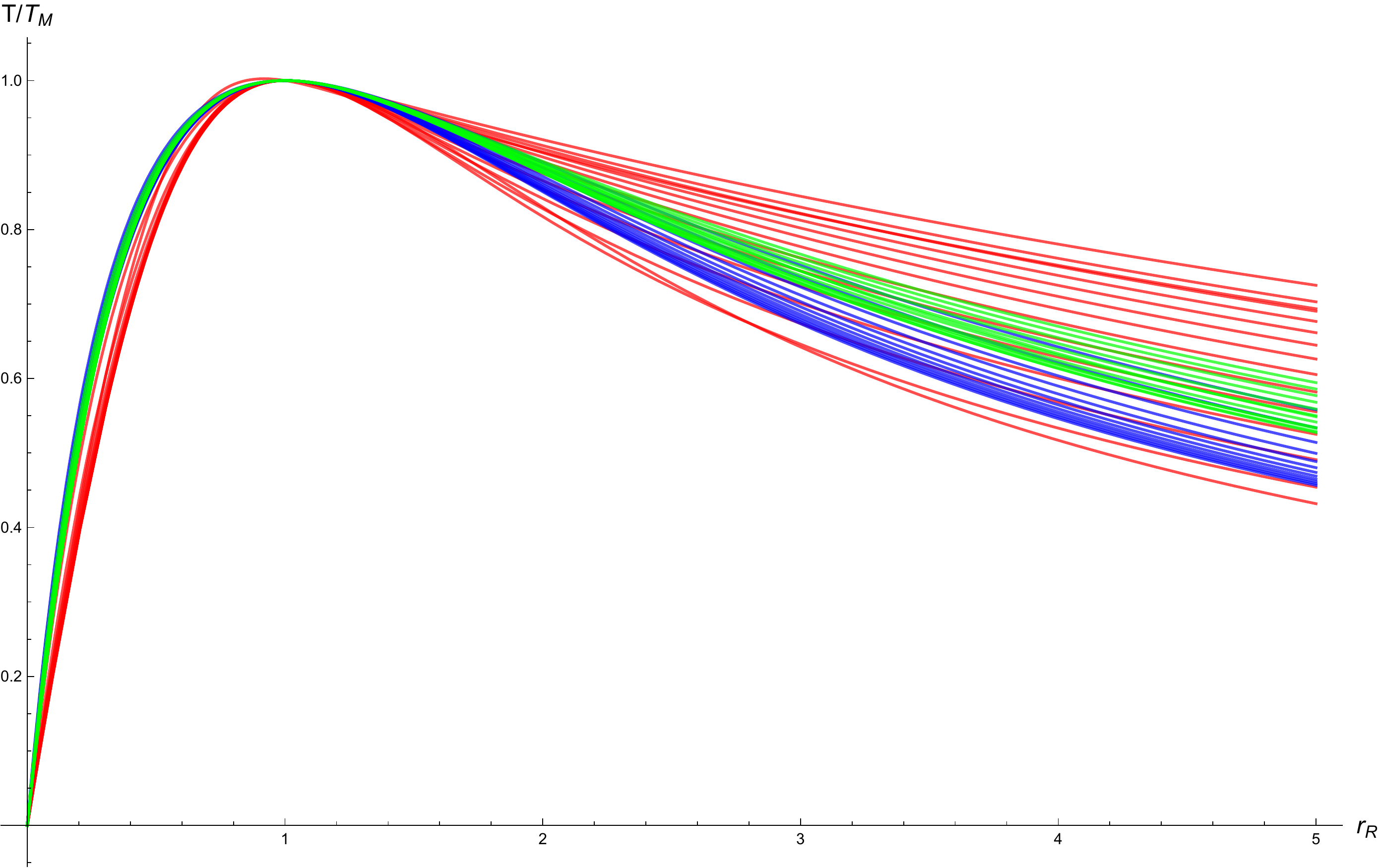}
    \caption{Temperature profiles for the densities \eqref{densities} with $ q \leq 15$ where $g_1$ are the red lines, $g_2$ are the blue lines and $g_3$ are the green lines---they mostly overlap. The plots have been rescaled so the maximal temperature is reached at the same point. The unscaled value of the maximal temperature in the considered cases is between $0.02\lambda^{-1}-0.1\lambda^{-1}$, where $\lambda^{-1}$ is of the order of the Planck temperature.}
   \label{figure temp}
\end{figure}

Also, since the derivative of $f(r)$ is finite everywhere, as is obvious from the regularity of $\rho(r)$ and the form of the solution \eqref{f int}, we see that the infinite Hawking temperature is avoided. Actually, in cases studied in \cite{Kovacik:2021qms}, the maximal temperature is an order or two below the Planck temperature, $T \sim \lambda^{-1}$, see Figure \ref{figure temp}.

The existence of black hole remnants have been derived from very few assumptions: having monotonically decreasing regular matter density $\rho(r)$ and Schwarzschild-like form of the solution. The microscopic black holes, before freezing into remnants, radiate with near-Planck temperatures; steepness of their temperature profile for the case of $\tilde{\rho}(r) \sim e^{-\left(r/\lambda\right)}$ has been derived in \cite{Kovacik:2017tlg}. The temperature reached after adding a small mass $\delta m$ to the minimal mass $m_0$ is
\begin{equation}
T(m_0 + \delta m) \stackrel{.}{=} \frac{\sqrt{\delta m / \lambda }}{41.01 } \lambda^{-1}.
\end{equation}

Parameters of these remnants, such as the mass or the cross-section, are naturally close to the Plank units. Typical temperature profiles are shown in Figure \ref{figure temp}.

This leads to two possible observational consequences. The first is that such black hole remnants are a good dark matter candidate \cite{Chen:2002tu}. If the minimal scale is the Planck length, then their cross-section is of the order of $10^{-70}m^2$---this makes them nearly impossible to interact with. However, with the mass of the order of the Planck mass, only a relatively small number of them would be needed to explain the observed amount of the dark matter density, approximately of the order of $n_{\mbox{\scriptsize{mbh}}} \sim 10^{-20}m^{-3}$ \cite{Kovacik:2017tlg}. If this is indeed the case, it could be difficult to verify as there is no straightforward way of detecting them. We will return to this question in the next section. 

The second possible observational consequence is the production of GRB during the final stage of the evaporation process. Contrary to the ordinary black holes, in the case of regular black holes is the infinite temperature avoided. There are very few quanta radiated during the peak emission. To put this into perspective, the Large hadron collider currently operates at energies $E \approx 10^{-15} E_p$, where $E_p$ is the Planck energy. The GRB with highest ever detected energy have reached and even surpassed this energy scale \cite{deUgartePostigo:2019pbh, Amenomori:2019rjd} and the cosmic ray with the largest energy had $E \approx 10^{-9} E_P$, \cite{Bird:1994uy, Waxman:1995vg}. Also, protons with $E > 10^{-9} E_P$ are prevented from traveling galactic distances by Greisen-Zatsepin-Kuz'min limit \cite{Greisen,Zatsepin}. In the studied models of regular black holes, the energy of Hawking-radiated particles peaked around $10^{-2} E_p$.  

Lack of observational evidence of the Hawking radiation from miscroscopic black holes does not disprove their existence but only puts some constraints on their mass spectrum \cite{Carr:2020gox}. However, only a small number, on the order of $10-100$, are radiated during the peak emission. Most of the energy is radiated at energy scales that overlap with more conventional astrophysical mechanisms. Also, the microscopic black holes could have finished the evaporation process during the earliest moments of the universe when the radiated energy would have been quickly absorbed. The search for a signal from evaporating primordial black holes is still ongoing \cite{Lopez-Coto:2021lxh, Ukwatta:2015iba, Berger:2013jza}.

Lacking observational signals, the other option is to try to connect the observed amount of dark matter with the theoretical predictions of primordial black hole production. However, their production could have taken place during the period of inflation, physics of which we only begin to understand \cite{Carr:2009jm}. The total number of black holes produced depends on the details of the inflation field \cite{Linde:2012bt, Solbi:2021rse, Martin:2019nuw,Garcia-Bellido:2017mdw}---another aspect that is still not perfectly understood. However, if at some point in the future the details of the inflation process will be known, we can use them to calculate the expected number of black hole remnants and compare them to the observed dark matter density; keeping in mind that the Planckian remnants could contribute only partially to the overall dark matter density. 

\section{Radiation recoil effect}

Recently, an important piece of the puzzle was added to this picture. In \cite{Kovacik:2021qms}, it was observed that the last fraction of the mass of microscopic black holes is radiated in a relatively small amount of quanta. From the point of having maximal temperature to the point of zero temperature, the black hole radiates approximately $10-100$ particles, this estimate is  only slightly model-dependent. 

The radiated particles have large momenta and random directions. Radiating each of them gives the black hole an opposite momentum kick. As a result, during the final moments of the evaporation process the microscopic black hole makes a random walk in the momentum space. Therefore, black hole remnants are formed with considerable velocity, typically close to $0.1$ of the speed light.

This effect was tested in \cite{Kovacik:2021qms} on three (infinite) classes of matter distributions
\begin{eqnarray} \label{dens} \nonumber
\tilde{\rho}_1(r; q) &=& n_1 e^{-\left(r/\lambda\right)^q},\\
\tilde{\rho}_2(r; q) &=& n_2 \left(1+\left(r/\lambda\right)\right)^{-q},\\ \nonumber
\tilde{\rho}_3(r; q) &=& n_3 \left(1+\left(r/\lambda \right)^q\right)^{-1}.
\end{eqnarray}

Solutions to the corresponding Einstein field equations are, using the parametrisation $f_i(r;m,q) = -1 + \frac{2m}{r} g_i(r;q)$, following
\begin{eqnarray}  \label{densities} \nonumber
g_1(r;q) &=& \frac{3\ \left( \Gamma\left(3\ q^{-1}\right) - \Gamma\left(3\ q^{-1},r^q\right)\right)}{q \ \Gamma\left(1 + 3\ q^{-1}\right)} ,\\  
g_2(r;q) &=&1 - \frac{ \left(2 + \left( q-1\right) r \left(2 + \left(q-2\right) r\right)\right)}{2\left(1 + r\right)^{q-1}},\\ \nonumber
g_3(r;q) &=& \frac{q\ r^3 \ \setlength\arraycolsep{1pt} {}_2 F_1 \left(1,3\ q^{-1},1+3\ q^{-1},-r^q\right) \sin\left(3 \pi\ q^{-1}\right) }{3\pi} ,
\end{eqnarray}
where $\Gamma(a,b)$ is the upper incomplete Gamma function and $\setlength\arraycolsep{1pt}
{}_2 F_1$ is the hypergeometric function. 

\begin{figure}[h] 
\centering
  \includegraphics[width=1.0\textwidth]{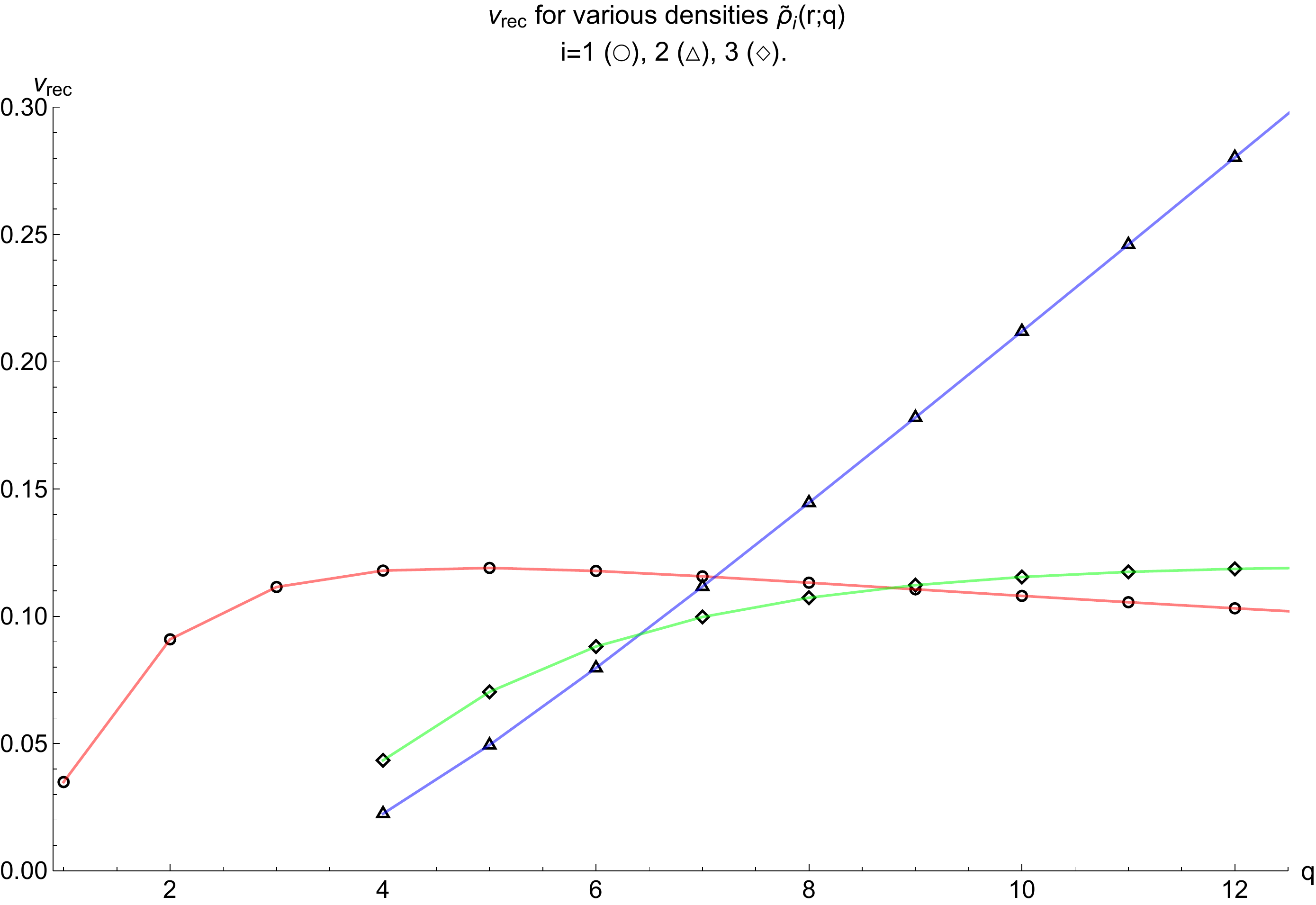}
    \caption{Recoil velocities for the densities \eqref{densities} with $ q \le 12$ expressed in units where $c=1$. As in the previous plot: the result for $g_1$ is shown by the red line, $g_2$ by the blue line, and $g_3$ by the green line. The values were computer for integer values of $q$, the lines are only joining the points to help to see the overall trend. As discussed in \cite{Kovacik:2021qms}, the results for $g_3$ can be trusted only for $q \approx 10$ as above this point the description breaks down, the mass $\Delta m$ is radiated away in a single quantum and the description needs to be refined.}
   \label{figure vr}
\end{figure}

Note that the cases of $\tilde{\rho}_1(r; q)$ with $q=1,2,3$ have been studied before, $q=1,2$ in the context of quantum spaces and $q=3$ in the context of vacuum polarisation and some others were studied in the context of dark matter halos \cite{Batic:2021mkm}. The recoil velocity for each of those cases are shown in Figure \ref{figure vr}, its value has been derived in \cite{Kovacik:2021qms}
\begin{equation} \label{rec}
v_{\mbox{\small{rec}}} \approx  \frac{\Delta m}{m_0 \sqrt{N_q}},
\end{equation}

where $\Delta m = m_{\mbox{\scriptsize{Tmax}}}-m_0$ is the mass difference between black holes with maximal and zero temperature and $N_q$ is the number of quanta radiated during the transition between them.

Such velocities would make microscopic black holes incompatible with the current picture of \textit{cold} dark matter. How to avoid this issue? The study \cite{Lehmann:2021ijf} suggested that if the evaporation finished a long time ago–––as also corroborated from the theory of big bang nucleosynthesis that is sensitive to possible energy injections---the black hole remnants would have been slowed down by the expansion of space. In \cite{Kovacik:2021qms}, it was calculated that for this the happen, the primordial black holes would need to have formed before the universe was $ 10^{-28} \mbox{s}$ old, which is shortly after the end of the inflation era.

Details of the Hawking radiation and therefore also the recoil effect are not well-understood at the moment and have been investigated in a different context in \cite{Frolov:2002gf, Dai:2007ki}.

\section{Conclusion}

Despite the fact that we expect the effects of quantum gravity to be dominant at the Planck energy scale, various mechanisms make this theory relevant well below this scale. The exact formulation of quantum gravity is not understood at the moment, however, some of its properties are rather general and therefore we can hope to make some phenomenological predictions without knowing the exact details.

We have discussed the effect of the quantum structure of space on microscopic black holes that have been proposed as possible dark matter candidates. Even though there is some effort to observe GRB signal from the final moments of evaporation, there are two arguments why the evaporation process has to have ended at very early times. The first is the agreement between the theory and observation of the Big bang nucleosynthesis process which is sensitive to energy injection \cite{Kohri:1999ex, Acharya:2020jbv}. The second is the recoil effect which would make the late-evaporation black holes incompatible with the current cold dark matter hypothesis.

This estimate is in line with works investigating the formation of black holes during the inflation epoch or during the reheating phase when the inflation ceased \cite{Carr:2018nkm,Padilla:2021zgm}. However, there is yet another option---could the Planck remnants have been formed during the Planck epoch? If this is the case, there is no obvious way how to compute their properties, such as their mass distribution, given our current understanding.

One thing needs to be stressed, there are in principle two types of black primordial black holes that have been conjectured to be dark matter constituents. One of them are the Planck remnants, which, as we have discussed in this paper, finished their evaporation process at a very early time, \cite{Kovacik:2021qms,Lehmann:2021ijf, Kohri:1999ex,Acharya:2020jbv}. The other class consists of heavy black holes for which the evaporation is not so important and perhaps outpaced by accretion \cite{Mukherjee:2021itf}. Our discussion only concerns the first type, the Planck-mass black holes.

The black hole models were derived from the mass density functions $\rho(r)$ which were either defined by hand or derived from simple models of quantum spaces. Currently, the best candidate for the quantum theory of gravity is the M-theory. It would be therefore interesting to use this theory to derive $\rho(r)$ and to study semi-classical consequences of the M-theory. Another option is to compare the presented results with the quantum loop gravity---another prominent quantum gravity hypothesis---description of black holes \cite{Perez:2017cmj}. If not for anything else then just to understand better the generality of the results presented here.

\section*{Acknowledgement}
The author wish to thank the Corfu Summer Institute for its hospitality. This research was supported by VEGA 1/0703/20 and the MUNI Award for Science and Humanities funded by the Grant Agency of Masaryk University. Valuable comments from V. Balek, J. Tekel and N. Werner were greatly appreciated.

\end{document}